\documentclass[aps,prb,preprint,superscriptaddress,showpacs]{revtex4-2}

\usepackage{graphicx}
\usepackage[english]{babel}
\usepackage{amsmath,amssymb,amsfonts,latexsym,amsbsy}
\usepackage{gensymb}
\usepackage[colorlinks=true,linkcolor=blue,citecolor=blue,urlcolor=blue]{hyperref}
\usepackage{xcolor}
\usepackage{lineno}
\usepackage{upgreek}
\usepackage{tabularx}
\usepackage[normalem]{ulem}

\begin{document}


\title{Decoupling between propagating acoustic waves and two-level systems in hydrogenated amorphous silicon}


\author{M. Molina-Ruiz}
\email[Corresponding author: ]{manelmolinaruiz@gmail.com}
\affiliation{Department of Physics, University of California Berkeley, Berkeley, California 94720, USA}

\author{H. C. Jacks}
\altaffiliation{Present address: Department of Physics, California Polytechnic University, San Luis Obispo, California 93407, USA}
\affiliation{Department of Physics, University of California Berkeley, Berkeley, California 94720, USA}

\author{D. R. Queen}
\altaffiliation{Present address: Northrop Grumman Corp., Linthicum, Maryland 21090, USA}
\affiliation{Department of Physics, University of California Berkeley, Berkeley, California 94720, USA}

\author{T. H. Metcalf}
\affiliation{Naval Research Laboratory, Washington, DC 20375, USA}

\author{X. Liu}
\affiliation{Naval Research Laboratory, Washington, DC 20375, USA}

\author{F. Hellman}
\affiliation{Department of Physics, University of California Berkeley, Berkeley, California 94720, USA}


\date{\today}

\begin{abstract}
Specific heat measurements of hydrogenated amorphous silicon prepared by hot-wire chemical vapor deposition show a large density of two-level systems at low temperature. Annealing at 200 \degree C, well below the growth temperature, does not significantly affect the already-low internal friction or the sound velocity, but irreversibly reduces the non-Debye specific heat by an order of magnitude at 2 K, indicating a large reduction of the density of two-level systems. Comparison of the specific heat to the internal friction suggests that the two-level systems are uncharacteristically decoupled from acoustic waves, both before and after annealing. Analysis yields an anomalously low value of the coupling constant, which increases upon annealing but still remains anomalously low. The results suggest that the coupling constant value is lowered by the presence of hydrogen.
\end{abstract}


\maketitle
Amorphous solids show anomalous thermal, elastic and dielectric properties at low temperatures. The observation of anomalous thermal properties, through thermal conductivity and specific heat measurements~\cite{Zeller1971}, led in 1972 to development of the standard tunneling model (STM)~\cite{Phillips1972,Anderson1972}, in which atoms or groups of atoms tunnel between nearly degenerate configurations with a distribution of asymmetries, $\Delta$, that are separated by energy barriers on the order of tens of Kelvin with a distribution of tunneling parameter, $\lambda$, resulting in tunneling-induced states with splittings $<$ 1 K. The simplest description of these configurations and the resulting states are two-level systems (TLSs). The STM was further developed during the following years, providing an explanation for various anomalous low temperature properties~\cite{Hunklinger1976, Halperin1976, Black1977, Black1978, Pohl1981, Zimmerman1981, Phillips1987}. The specific structures that enable the TLSs are generally not known, but have been proposed for some specific materials, such as silica~\cite{Damart2018}. Recently, we suggested that TLSs in hydrogenated amorphous silicon ($a$-Si:H) are primarily due to clustered atomic H in low density regions~\cite{Molina-Ruiz2020}.

The STM assumes that the asymmetry between states, $\Delta$, is small, and that the tunneling parameter, $\lambda$, is uniformly distributed over a range of values, which allows the TLS distribution function, $P(\Delta,\lambda)$, to be written as $P(\Delta,\lambda) \approx \bar{P}$, where $\bar{P}$ is the TLS density. Under this assumption, the specific heat, $C_P(T)$, of an ensemble of TLS has a linear term, $c_1$, associated with the density of TLS, $n_0$, that equilibrate with the phonon bath on the time scale of the measurement by
\begin{equation}
\text{\small{$C_{P}(T)=c_1T=\frac{\pi^2}{6}\frac{k_B^2N_A}{n_{at}}n_0T$}}
\label{equation1}
\end{equation}
where $n_{at}$ is the total atomic density, $k_B$ Boltzmann's constant, and $N_A$ Avogadro's number. The specific heat derived TLS density, $n_0$, is proportional to the TLS density, $\bar{P}$, through a relationship first proposed by Black and Halperin~\cite{Black1977,Black1978}, and later experimentally verified~\cite{Meissner1981ExperimentalSilica,Loponen1982}, which establishes
\begin{equation}
\text{\small{$n_0=\frac{1}{2}\bar{P}\ln{\left(\frac{4t}{\tau_{min}}\right)}$}}
\label{equation2}
\end{equation}
where $t$ is the measurement time ($\approx10$ ms for the nanocalorimetric system used in the present work at low T)~\cite{Denlinger1994}, and $\tau_{min}$ is the TLS minimum relaxation time. At low temperatures, typically below 10 K, the main TLS relaxation mechanism is the one-phonon process, in which $\tau_{min}$ is minimized when the energy difference between two TLS states $E=\sqrt{\Delta^2 + \Delta_0^2}=\Delta_0$, where $\Delta_0$ is the TLS tunnel splitting. This condition implies that the minimum relaxation time is achieved by symmetric TLSs ($\Delta=0$). In dielectric glasses, the main contribution to specific heat and mechanical loss comes from TLSs with energy splitting $E = k_B T$, 
and their minimum relaxation time is given by~\cite{Jackle1972, Esquinazi1998, Enss2005}
\begin{equation}
    \text{\small{$\tau_{min} = \left(\frac{\gamma_\ell^2}{v_\ell^5} + 2\frac{\gamma_t^2}{v_t^5}\right)^{-1} \frac{\pi\rho\hbar^4}{k_B^3} T^{-3} = a T^{-3}$}}
    \label{equation7}
\end{equation}
where subscripts $\ell$ and $t$ denote the longitudinal or transverse wave polarization, respectively, which from now on and as a generalization are denoted by the subscript $\alpha$. $\gamma_\alpha$ is the coupling constant between TLS and phonons, $v_\alpha$ the sound velocity, $\rho$ the mass density, and $T$ the system temperature. Typically, $a \approx 10^{-8}$ s\,K$^3$~\cite{Jackle1972, Esquinazi1992}, and therefore $\tau_{min} \approx 1$ ns at 2 K. 
Note that because of the logarithmic time-dependence in Eq.~\ref{equation2}, the ratio $n_0 / \bar{P}$ depends only weakly on the measurement time, $t$, and on the TLS minimum relaxation time, $\tau_{min}$, such that $n_0 / \bar{P} \approx 10$ for essentially all experimentally realized measurements~\cite{Grace1989}.

The interaction between phonons and TLS at low temperature leads to energy loss, which may occur by resonant or relaxation mechanisms. At temperatures above $\sim10$ K, losses arise from thermal activation over the energy barriers separating the states. At the experimental temperatures and excitation frequencies, $\omega$, used in the present work, relaxation dominates over resonant mechanisms through TLSs with $\omega \tau_{min} = 1$. The interaction between a phonon and a TLS is given by the deformation potential, $\gamma_{i,\alpha} = \partial\Delta_i/2\partial{u_\alpha}$, where $u_\alpha$ is the strain tensor. $\gamma_{i,\alpha}$ thus depends on the specific TLS. 
In the limit assumed by the STM, a single value of the deformation potential is used for all TLSs, the coupling constant $\gamma_\alpha$. In that limit and in the relaxation regime, the mechanical loss, $Q_\alpha^{-1}(T)$, is temperature-independent, typically from 0.1 to 10 K~\cite{Hunklinger1976}, and is given by
\begin{equation}
    \text{\small{$Q_{0,\alpha}^{-1}=\frac{\pi}{2}\frac{\bar{P}{\gamma_\alpha}^2}{\rho{v_\alpha}^2}$}}
\label{equation4}
\end{equation}
Experimentally, $\gamma_\alpha$ can be determined combining acoustic dispersion and attenuation measurements~\cite{Jackle1976}, or specific heat and acoustic measurements~\cite{Duquesne1983}. Mechanical loss, $Q_\alpha^{-1}(T)$, measurements report the imaginary part of the elastic constant, the so-called anelastic energy loss. In amorphous solids, other contributions to $Q_\alpha^{-1}(T)$ are negligible in this temperature range. Surprisingly, $Q_{0,\alpha}^{-1}$ ranges from 10$^{-4}$ to 10$^{-3}$ for most glasses and varies little with differing chemical compositions~\cite{Pohl2002}. This ``universal'' glassy range is not a consequence of the STM, and has been long considered one of the great mysteries of glasses. Relatively recent explanations of this effect center on the concept of self-interacting TLS, which renormalize the TLS density and yield approximately the observed universal number~\cite{Yu1988,Carruzzo2020}.

The relative change with temperature of the sound velocity, $\Delta{v_\alpha}/v_{0,\alpha}$, reports the real part of the elastic constant, and is measured from the relative variation of the resonant frequency of the oscillator with temperature, $\Delta{f}/f_0$, with $f_0$ being the frequency at a reference temperature $T_0$. Elastic properties measurements have shown that the thermally-activated relaxation rate of the TLS cause a reduction in $\Delta{v_\alpha}/v_{0,\alpha}$ proportional to temperature,
\begin{equation}
\text{\small{$\frac{\Delta{v_\alpha}}{v_{0,\alpha}}=-\beta_\alpha(T-T_0)$}}
\label{equation5}
\end{equation}
where $v_{0,\alpha}$ is the sound velocity at an arbitrary reference temperature $T_0$. Eq.~\ref{equation5} is not predicted by the STM, but $\beta_\alpha$ has been experimentally found to be proportional to $Q_{0,\alpha}^{-1}$ for a large number of systems, including amorphous solids, quasicrystals and disordered crystals. Specifically, the average proportionality for all the studied systems yields $\beta_t / Q_{0,t}^{-1} = 0.5$ K$^{-1}$~\cite{White1996}. The results presented in this work are obtained from transverse modes, so from now on, and unless noted, we always refer to transverse components and omit the wave polarization notation.

It has been shown that vapor deposited films grown under specific conditions surpass constraints that limit the properties of glasses obtained from the liquid phase, producing ultrastable glasses, i.e., glasses with higher glass transition temperature and atomic density~\cite{Ediger2014, Perez-Castaneda2014, Ediger2017}. A correlation between atomic density and TLS density, derived from both specific heat and mechanical loss (extending far below the ``universal'' glassy range), in $a$-Si films grown under conditions suggesting ultrastability has been shown~\cite{Queen2013, Liu2014, Queen2015Two-levelSilicon}. $a$-Si:H was the first material to report $Q_0^{-1}$ values below the ``universal'' glassy range, which was initially suggested to be due to the presence of hydrogen~\cite{Liu1997a, Liu1998}, but was subsequently found as well in $a$-Si (without H) grown at elevated temperature~\cite{Liu2014, Queen2015Two-levelSilicon}. Recently, the TLS density of $a$-Si:H films grown under different conditions has been studied by specific heat, $C_P(T)$, measurements~\cite{Molina-Ruiz2020}. The aim of this work is to compare $n_0$ and ${Q_0}^{-1}$ measured on equivalent samples of $a$-Si:H in the as-prepared and annealed states, where it is known that annealing allows hydrogen to redistribute while preserving the silicon network~\cite{Molina-Ruiz2020}, and thus obtain a better understanding of the STM.

We report specific heat, $C_P(T)$, mechanical loss, $Q^{-1}(T)$, longitudinal sound velocity, $v_\ell$, and relative change of the sound velocity, $\Delta{v}/v_0$, measurements on device quality $a$-Si:H films prepared by hot-wire chemical vapor deposition (HWCVD), at a growth rate of 2 nm/s, and at a substrate temperature $\textrm{T}_\textrm{S}$ = 370 \degree C, yielding films with 7\% atomic hydrogen (7 at.\% H). This preparation yields device-quality material with low dangling bond density~\cite{Mahan1991}. Growth, infrared spectroscopy and time-domain thermoreflectance characterization of similar HWCVD films were discussed elsewhere~\cite{Mahan1991,Yang2010}. The total atomic density (Si + H atoms) was determined by Rutherford backscattering spectrometry, in combination with a KLA-Tencor Alpha-Step IQ profilometer used to measure the films' thickness ($\sim85$ nm). These data show that the total atomic density depends non-monotonically on growth temperature and at.\% H. $v_\ell$ was measured at room temperature by a picosecond ultrasonic pump-probe technique~\cite{Lee2005}, yielding 7.7 km/s. Oxygen resonant scattering was used to look for oxygen content and none was detected ($<$ 1 at.\%). Films were measured in their as-deposited state, then annealed in vacuum at 200 \degree C for 10$^4$ seconds, and measured again. 
Hydrogen forward scattering was used to measure H content, and in particular it proved that annealing did not remove any hydrogen from the films. NMR measurements show that annealing causes H to disperse more uniformly, primarily remaining in isolated Si--H bonds and sometimes forming H$_2$ molecules~\cite{Beyer1983,Baugh1999}. Clustering of H$_2$ can occur and cause bubbles, but this was not observed in the present films under the annealing conditions used.

$C_P(T)$ was measured, both before and after annealing, from 2 to 300 K using the small $\Delta{T}$ relaxation method~\cite{Denlinger1994} and a microfabricated thin-film nanocalorimeter~\cite{Queen2009}. $Q^{-1}(T)$ and $\Delta{v}/v_0$ measurements were made from 0.3 to 100 K using the anti-symmetric torsional mode of a double paddle oscillator at $\omega$ = 2$\pi$ 5431 Hz~\cite{White1995}. We note that the interaction between TLS and phonons is dominated by relaxation processes, for both $C_P(T)$ and $Q^{-1}(T)$, in these experimental conditions. An extended set of specific heat measurements and their analysis are reported elsewhere~\cite{Molina-Ruiz2020}, showing the effect of growth temperature, annealing, and H content on the total specific heat atomic density, sound velocity, and TLS density, $n_0$, for a range of HWCVD $a$-Si:H thin films.

\begin{figure}
\includegraphics{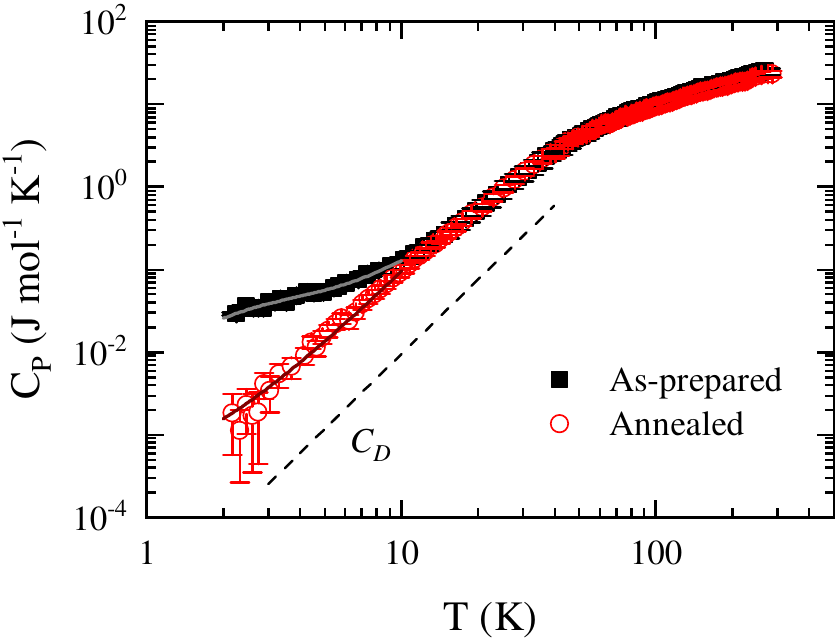}
\caption{\label{figure1}Specific heat of $a$-Si:H film grown at $\textrm{T}_\textrm{S}$ = 370 \degree C (7 at.\% H) in the as-prepared and annealed states. The Debye specific heat (dashed line), $C_D = c_DT^3$, is calculated from the measured sound velocity $v_\ell$. The solid grey and brown lines show the fitting to the as-prepared and annealed states, respectively, using Eq.~\ref{equation6}.}
\end{figure}

Specific heat data on this 7 at.\% H sample are shown in Figure~\ref{figure1}. The sound velocity and hence the Debye specific heat, $C_D$, due to the phonon contribution did not change on annealing and is shown with a dashed line. $C_D$ was calculated from the measured atomic density and longitudinal sound velocities as $C_D = (12\pi^4/5) N_Ak_B {(T/T_D)}^3$, where $T_D = (\hbar/k_B) (6\pi^2n_{at} [1/(3{v_l}^3) + 2/(3{v_t}^3)]^{-1})^{1/3}$ is the Debye temperature. Transverse sound velocity was calculated using the relationship $v_t$ = (0.56 $\pm$ 0.05) $v_\ell$, established between longitudinal and transverse modes for amorphous materials~\cite{Berret1988}, including our own work on $a$-Si~\cite{Queen2015Two-levelSilicon}.

The specific heat is significantly larger than the Debye specific heat, as is commonly found for amorphous materials. Data for the as-prepared and annealed states match above 10 K, making clear that vibrational modes are unchanged and the sample has not been modified by annealing, except for its H distribution and TLS parameters. Below 10 K, the data for the as-prepared sample is significantly larger than that of the annealed sample and shows a sub-linear temperature dependence, which must represent the high temperature side of a peak in order to have non-divergent entropy as T $\rightarrow$ 0.  The simplest and most likely model for this peak is a Schottky anomaly~\cite{Molina-Ruiz2020}. $C_P(T)$ is therefore fit to the following form:
\begin{equation}
\text{\small{$C_P(T)=C_{Sch}(T)+c_1T+c_3T^3$}}
\label{equation6}
\end{equation}
where $C_{Sch}(T)$ is the Schottky specific heat given by $C_{Sch}(T) = Nk_B {\left(\frac{\Delta_0}{k_B T}\right)}^2 \frac{\exp{(\Delta_0/k_B T)}}{[1+\exp{(\Delta_0/k_B T)}]^2}$, with $N$ the number of states with energy splitting $\Delta_0$, $c_1$ is given by Eq.~\ref{equation1}, and $c_3 = c_D + c_{ex}$, with $c_D$ the Debye specific heat coefficient and $c_{ex}$ an excess cubic coefficient that is not predicted in the STM but has also been reported for $a$-Si thin films~\cite{Queen2013}, and is discussed at length in Ref.~\cite{Molina-Ruiz2020}, including a discussion of the form of this fit.

As discussed in Ref.~\citep{Molina-Ruiz2020}, after annealing, $C_P(T)$ below 10 K is significantly reduced but remains larger than $c_D$ due to both $c_1$ and $c_{ex}$, and, in particular, the Schottky term is gone. We note that the low T $C_P(T)$ of $a$-Si:H and $a$-Si does not depend on magnetic field, where we observed variations $<1\%$ in fields up to 8 T (see Fig. 7.7 in Ref.~\cite{Queen2011}). These results confirm that electronic states are not responsible for the excess low T $C_P(T)$.

The TLS density, $n_0$, is calculated from the data in Figure~\ref{figure1} using Eq.~\ref{equation6}, and is reported in Table~\ref{table1}. The fits for $n_0$ and $c_{ex}$ are robust to details of the Schottky fit needed to explain the as-deposited data (e.g. multiple energy levels, varying $N$ while fixing $\Delta_0$, degeneracy, etc.) as discussed in Ref.~\citep{Molina-Ruiz2020}, which showed and discussed the specific heat of several samples grown at different temperatures and exhibiting different H content. We note here only that the lowest value of $n_0$ is obtained from the sample reported in this work ($\text{T}_\text{S} = 370$ \degree C and 7 at.\% H) both before and after annealing.

\begin{table}
\caption{\label{table1}Sample state: as-prepared or annealed at $\textrm{T}_\textrm{A} = 200$ \degree C, Schottky specific heat, $C_{Sch}$, at 2.4 K, TLS density, $n_0$, calculated from Eq.~\ref{equation1}, slope of the relative change of the sound velocity with temperature, $\beta$, obtained from Eq.~\ref{equation5}, and temperature-independent mechanical loss, $Q_0^{-1}$, at 1 K.}
    \begin{tabularx}{0.96\textwidth}{>{\centering\arraybackslash}X >{\centering\arraybackslash}X >{\centering\arraybackslash}X >{\centering\arraybackslash}X >{\centering\arraybackslash}X}
    \hline \hline
	state & $C_{Sch}$ & $n_0$ & $\beta$ & $Q_0^{-1}$ \\
	 & $\times10^{-3}$ & $\times10^{46}$ & $\times10^{-6}$ & $\times10^{-5}$ \\
	 & \small{(J\,mol$^{-1}$\,K$^{-1}$)} & \small{(J$^{-1}$m$^{-3}$)} & \small{(K$^{-1}$)} &  \\ \hline
	as-prep & 11.4$\pm$3.7 & 209$\pm$26 & 11.4$\pm$0.2 & 1.07$\pm$0.02 \\
    annealed & 0 & 29$\pm$3 & 12.2$\pm$0.5 & 1.15$\pm$0.02 \\
 \hline \hline
\end{tabularx}
\end{table}

\begin{figure}
\includegraphics{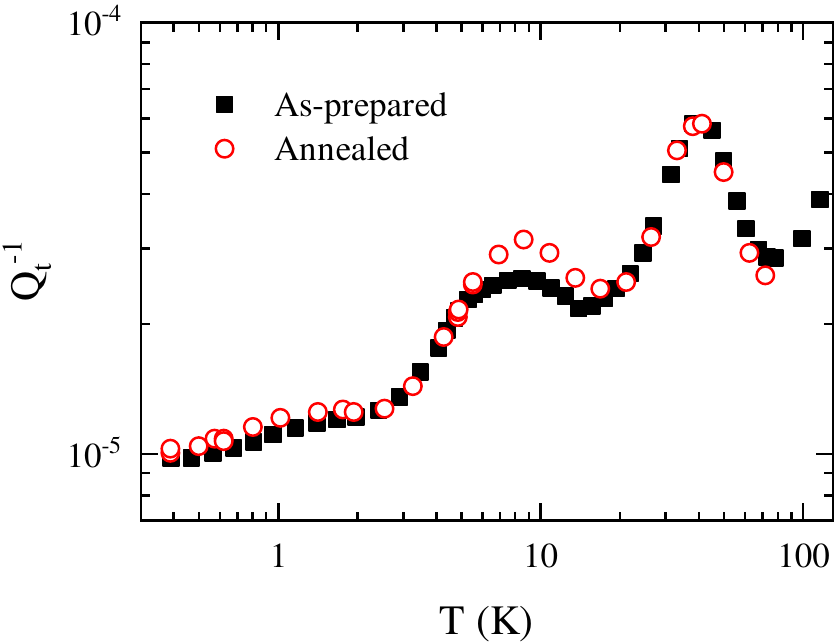}
\caption{\label{figure2}Mechanical loss, $Q^{-1}$, as a function of temperature, $T$, for an $a$-Si:H sample grown at 370 \degree C in its as-prepared (black closed squares) and annealed (red open circles) states.}
\end{figure}

Mechanical loss, $Q^{-1}(T)$, of $a$-Si:H in the as-prepared and annealed states was measured using ring-down measurements of a double paddle oscillator (DPO) before and after deposition of the a-Si:H sample, and also after annealing~\cite{White1995}. Results of $Q^{-1}(T)$ are shown in Fig.~\ref{figure2}. $Q_0^{-1} \sim 1 \times 10^{-5}$ for the as-deposited state, and increases $\sim8\%$ upon annealing. For other HWCVD $a$-Si:H, $Q_0^{-1}$ has been found to range from $4\times10^{-7}$ to $2\times10^{-5}$~\cite{Liu1998b}, consistent with our data; no explanation for this variation in $Q_0^{-1}$ has yet been suggested. Figure~\ref{figure2} also shows two peaks: at 6 K (8 K after annealing) and 40 K. The peak near 6 K is likely due to H$_2$ in a non-bulk environment; either caused by the liquid-solid transition of small clusters of H$_2$ in restricted geometries~\cite{Tell1983}, or to the occurrence of two-dimensional H$_2$~\cite{Liu1995}, which suppress the transition temperature. The lack of an H$_2$ triple point signature in both $C_P(T)$ and $Q^{-1}(T)$ suggests the latter possibility, specifically that H$_2$ clusters exist but are small enough that they do not behave as bulk H$_2$. After annealing at 200 \degree C, the 6 K peak intensity is enhanced by almost 30\% and becomes wider, which suggests either an increase in the H$_2$ content, or in the clustered/isolated H$_2$ ratio, consistent with literature on the effects of annealing below 200 \degree C~\cite{Beyer1983,Baugh1999}. The 40 K peak is not seen in any $a$-Si:H literature of which we are aware and is unchanged by annealing; it might be associated with contamination of some type.

\begin{figure}
\includegraphics{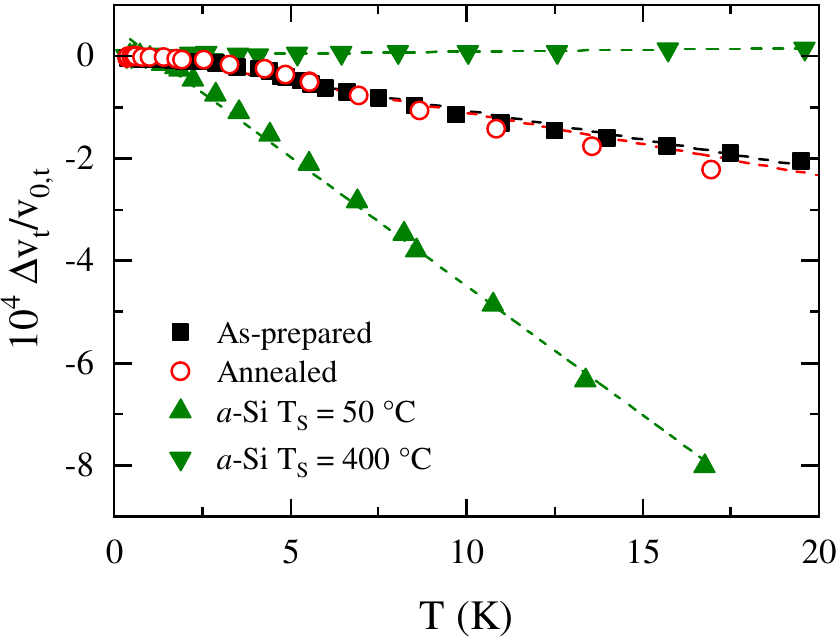}
\caption{\label{figure3}Relative change of the sound velocity, $\Delta{v_t}/v_{0,t}$, as a function of temperature, $T$, of $a$-Si:H in both as-prepared (black solid squares) and annealed (red open circles) states. For comparison, $\Delta{v_t}/v_{0,t}$ of $a$-Si samples grown at 50 \degree C (green up triangles) and 400 \degree C (green down triangles) are also shown from Ref.~\cite{Liu2014}. Dashed lines are linear fits showing $\beta$ (see Table~\ref{table1}).}
\end{figure}

Data on $\Delta{v}/v_0$ for the $a$-Si:H sample are shown in Fig.~\ref{figure3}. Again little change is observed between the as-prepared and annealed states, $\sim7\%$ increase in $\beta$, consistent with the 8\% increase seen above in $Q_0^{-1}$. We note the ratio between dispersion and attenuation properties $\beta / Q_0^{-1} = 1.1$  K$^{-1}$, which is more than two times the average experimental ratio of 0.5 K$^{-1}$ reported in previous works~\cite{White1996, Liu1998b}. For comparison, $\Delta{v}/v_0$ for $a$-Si is also shown in Fig.~\ref{figure3} for low and high growth temperatures~\cite{Liu2014}, which yields high and low TLS density, respectively, and is quantitatively consistent with the data for $a$-Si:H.

To go further with the analysis, we calculate $\bar{P}$ of $a$-Si:H from the specific heat values of $n_0$ using Eq.~\ref{equation2}, along with the experimental value of $t = 10$ ms, and $\tau_{min} \approx 1.25$ ns, obtained from Eq.~\ref{equation7} with $T = 2$ K, the lowest temperature of the specific heat measurements presented in this work. This approach yields $n_0 / \bar{P} \approx 9$, and therefore $\bar{P} \approx 2.4\times10^{47}$ J$^{-1}$m$^{-3}$ before annealing, and $\bar{P} \approx 3.4\times10^{46}$ J$^{-1}$m$^{-3}$ after. We note again that the logarithm in Eq.~\ref{equation2} makes the choice of $\tau_{min}$ not important (three orders of magnitude change in $\tau_{min}$ modifies $n_0/\bar{P}$ less than two-fold).

The experimental values of $Q_0^{-1}$ along with the values of $\bar{P}$ calculated above, and combined in Eq.~\ref{equation4}, yield $\gamma = 7$ meV and 19 meV for the as-prepared and annealed states, respectively. These results, however, lead to the non-intuitive conclusion that both $\bar{P}$ and $\gamma$ must change upon annealing in such a way that $Q_0^{-1}$ barely changes. The usual approach of assuming a value for $\gamma$, e.g. 1 eV, leads to a value of $\bar{P}$ from $Q_0^{-1}$ that is orders of magnitude less than that found from $C_P$ measurements, i.e., to a decoupling between propagating acoustic waves from TLSs, and thus not explaining why the large $\bar{P}$ found in heat capacity does not cause losses (an equivalent statement to the low $\gamma$ derived on comparing the two measurements).

Very few studies have been made that estimate $\gamma$~\cite{Duquesne1983, Duquesne1985UltrasonicCompounds, Bellessa1985, Berret1988, Morr1989, Fefferman2017}, which typically ranges from 0.1 to 1.5 eV, is considered a material constant, and has not been measured for $a$-Si:H. While the values of $\gamma$ reported in this work are unexpectedly low, there is nothing in the STM that prevents such low values. These low $\gamma$ values are on the order of those reported in H--Nb systems~\cite{Bellessa1985, Morr1989}, suggesting that H atoms might be responsible for the tunneling states in $a$-Si:H. The increased value of $\gamma$, obtained for the annealed state, and the fact that the sound velocity does not change, imply that upon annealing the minimum relaxation time decreases significantly (7-fold) since $\tau_{min} \propto v^5/\gamma^2$ (see Eq.~\ref{equation7}). It is remarkable that $Q_0^{-1}$ and $\beta$, both proportional to $\bar{P}\gamma^2$, increase very slightly after annealing, whereas $n_0$ and thus $\bar{P}$ decrease noticeably. This apparent contradiction is mathematically solved if $\gamma$ increases after annealing. We know that sound velocity does not change upon annealing, which suggests that the silicon structure is not altered. Additionally, the small increase of $Q_0^{-1}$ and $\Delta{v}/v_0$ may be caused by local structural changes when hydrogen diffuses. However, if the structural rearrangement of hydrogen atoms leads, as measured by $C_P(T)$, to a reduction of TLSs that turn out to be those with the lowest $\gamma_i$ values, then $\gamma$ should increase. This hypothesis explains why a reduction in $\bar{P}$ upon annealing would correlate with an increase in $\gamma$, or vice versa, in $a$-Si:H.

Previous work has found that large variations in specific heat are often not associated with changes in thermal conductivity~\cite{Hunklinger1975, Stephens1976}. A similar inconsistency is found in $C_P(T)$ and $Q^{-1}(T)$ measurements of CVD deposited amorphous silicon nitride, notably a material that contains of order 1 - 2 at.\% H~\cite{Zink2004,Liu2007}. An alternative approach to the decoupling between specific heat and acoustic properties results was proposed by Black~\cite{Black1978}, in which the samples may contain ``anomalous'' TLSs, besides ``standard'' TLSs. Black argued that anomalous TLSs relax fast enough to be measured by specific heat techniques, but too slow to be detected by acoustic properties techniques. In this approach, Black considered an additional TLS density for the anomalous states that would only contribute to the specific heat measurements. Despite the addition of anomalous TLSs, which in some extent improved the agreement between the STM and the experimental results, Black could not satisfactorily reconcile theory and results. We suggest that these apparent discrepancies are due to a misunderstanding of the coupling mechanisms between TLSs and propagating waves, i.e., the deformation potential and, particularly, that it varies from TLS to TLS, and hence to the value of the coupling constant to be used within the STM. We \textit{speculate} that if $\gamma_i$ is not the same for all TLSs within the system, then the average value of such a distribution, $\gamma$, may depend on the relaxation times probed. We note that $\tau$ ranges from $\tau_{min}$ to $\sim30$ $\upmu$s for $Q^{-1}(T)$, and from $\tau_{min}$ to $\sim10$ ms for $C_P(T)$ results presented in this work.

In conclusion, the TLS density measured by $C_P(T)$ in HWCVD $a$-Si:H is significantly larger than that measured by $Q^{-1}(T)$. Furthermore, annealing promotes hydrogen redistribution while the silicon network is preserved, which leads to a reduction by an order of magnitude of the TLS density from $C_P(T)$ while $Q_0^{-1}$ and $\beta$ slightly increase. The coupling constant of $a$-Si:H, derived from the STM prediction of $n_0/\bar{P} \approx 10$, yields $\gamma=7$ meV, and increases to $\gamma=19$ meV after annealing, suggesting not only that $\gamma$ is not a constant material quantity, but also that its value strongly depends on the microstructure. The low values of the coupling constant suggest that the TLSs that couple to applied elastic fields in $a$-Si:H are likely related to the presence of hydrogen within the amorphous silicon network. The discrepancy observed between $C_P(T)$ and $Q^{-1}(T)$ measurements shows that propagating acoustic waves are decoupled from TLSs with very low coupling constants.
\begin{acknowledgments}
We thank E. Iwaniczko, Q. Wang, and R. S. Crandall and the National Renewable Energy Laboratory for preparation of the $a$-Si:H films; G. Hohensee and D. G. Cahill for sound velocity and thermal conductivity measurements; and A. Fefferman for helpful discussions. The UCB portion of this work was supported by the NSF grants DMR-0907724, 1508828 and  1809498. Work performed at NRL was supported by the Office of Naval Research.
\end{acknowledgments}
%
\bibliography{references}
\end{document}